\documentstyle[12pt,psfig]{article}

\begin{document} 

\title{$\eta^{\prime}$ Production in Proton-Proton
Collisions near Threshold\thanks{\emph{Supported 
by Forschungszentrum J\"ulich}}}
\author{A. Sibirtsev\thanks{\emph{On leave from the Institute of 
Theoretical and Experimental Physics, 117259 Moscow, Russia.}} 
\ \ and  W. Cassing \\
Institut f\"ur Theoretische Physik, Universit\"at Giessen \\
D-35392 Giessen, Germany}
\date{ }
\maketitle
\begin{abstract}
Within  the one-pion exchange model we calculate
the near-threshold ${\eta}^{\prime}$ production from $pp$ collisions
including the final state interaction between the protons. Since the 
description of the data is quite well we conclude that $\eta, \rho$ or $
\omega$ exchange currents either play no role or cancel each other
to a large extent in this reaction.
\end{abstract}

Recently the COSY-11~\cite{Moskal} and 
SPES III~\cite{Hibou} Collaborations have measured the 
$\eta^{\prime}$ production in proton-proton collisions 
at excess energies $\epsilon =\sqrt{s}-2m_N-m_{{\eta}^{\prime}}$
$\leq$ 10 MeV. As found in Ref.~\cite{Hibou} the calculation for
the $pp \to pp \eta^{\prime}$ cross section within 
the one-pion-exchange model -- based on a
comparison between $\eta $ and ${\eta}^{\prime}$ production amplitudes -- 
underestimates the experimental data by up to a factor of 2, such that
the $\rho$ and other heavy-meson exchange diagrams~\cite{Germond} 
should contribute substantially for $\eta^{\prime}$ 
production. In this short note we will argue that this discrepancy 
vanishes when including the average amplitude 
$|M_{\pi N \rightarrow \eta^\prime N}|$ from experimental 
data while also taking into account the  interaction
between the protons in the final-channel.

Here we calculate the ${\eta}^{\prime}$ production within
the one-boson exchange model first neglecting the 
Final-State-Interaction (FSI).
For the one-pion exchange the $pp \to pp \eta^{\prime}$  production 
amplitude~\cite{SiCa} then reads
\begin{equation}
\label{amp1}
M = g_{NN\pi} \ F(t) \ \bar{u}(p_1){\gamma}_5 u(p_a) \
\frac{1}{t-\mu^2} \ M_{\pi^0 p \to p \eta^{\prime}},
\end{equation}
where $g_{NN\pi}=13.59$~\cite{Swart} is the $pp\pi^0$ coupling 
constant, $t=(p_a-p_1)^2$ is the squared 4-momentum transfer from 
the initial to the final proton, $\mu $ is the pion mass
and $F(t)$ is the form factor for the $NN\pi$ vertex 
\begin{equation}
F(t) = \frac{\Lambda^2-\mu^2 }{\Lambda^2 -t}
\end{equation}
with a cut-off parameter $\Lambda=1.3$~GeV~in line with 
Ref.~\cite{Machleid}. 
In principle, $\eta, \rho$ and $\omega$ exchanges should also contribute 
as suggested by the analysis in Ref.~\cite{Hibou}, however, there are no
data on the coupling of the ${\eta}^{\prime}$ to the baryonic resonances,
which are necessary to include the exchanges of other mesons 
in a reliable way.  

\begin{figure}
\psfig{file=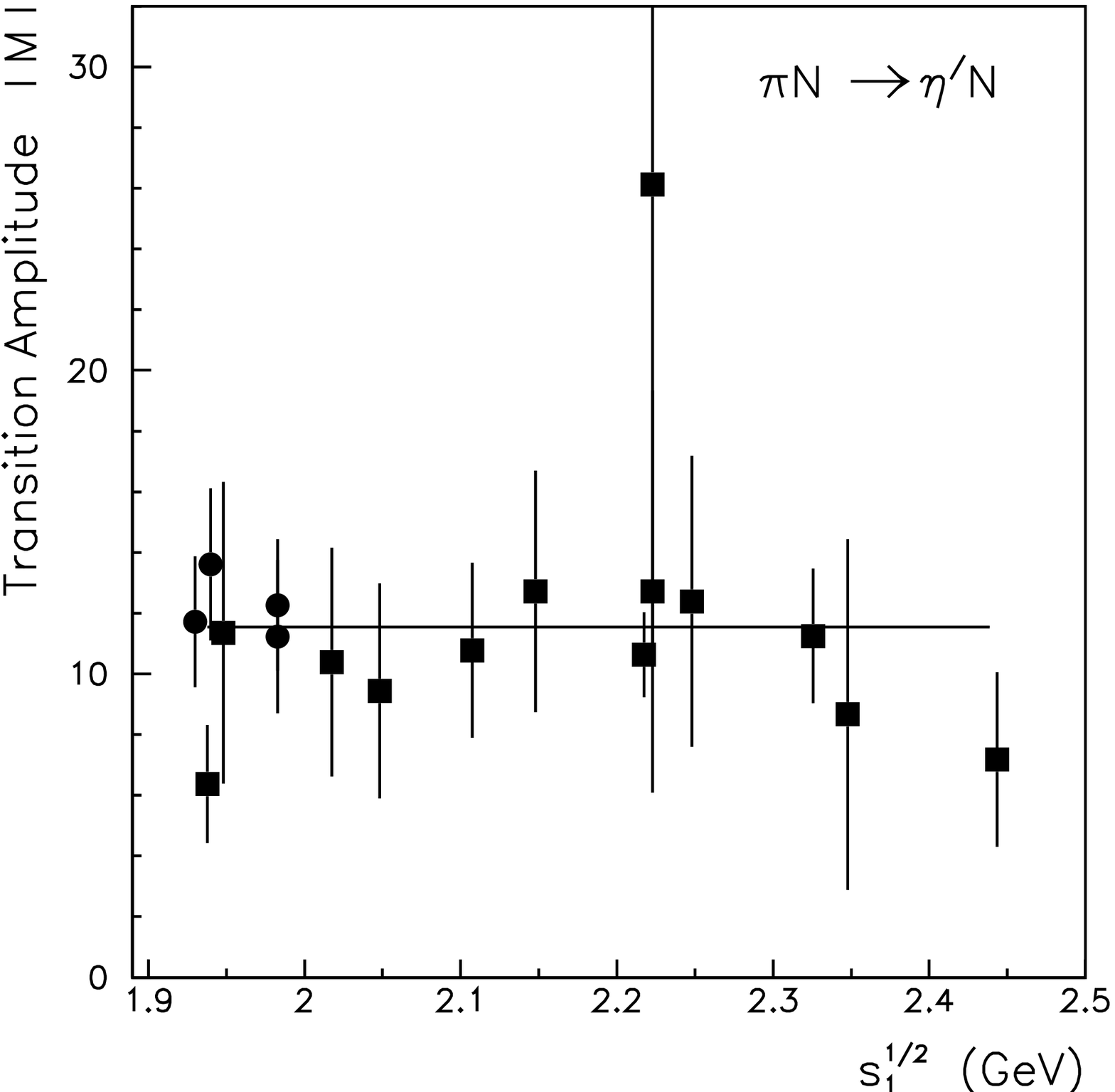,width=16cm}
\caption[]{The  amplitudes for the
$\pi^-p \to \eta^{\prime} n$ (circles) and
$\pi^+n \to \eta^{\prime} p$ reactions (squares)
as a function of the $\eta^{\prime} N $ invariant mass. The solid line
corresponds to the average amplitude used in our calculations. 
The symbols show the results extracted from the 
experimental data from Ref.~\protect\cite{LB} according to Eq. 
\protect\ref{amp2}).}
\label{fig1}
\end{figure}

In~(\ref{amp1}) $M_{\pi N \to N \eta^{\prime}} $ is
the amplitude for the ${\pi N \to N \eta^{\prime}}$ reaction,
which is related to the physical cross section as
\begin{equation}
\label{amp2}
|M_{\pi N \to \eta^{\prime} N}|^2 = 32 \pi  \ s_1 \  
\frac{q_{\pi}} {q_{\eta^{\prime}}} \
\sigma(\pi N \to \eta^{\prime} N) .
\end{equation}
Here $s_1$ is the squared invariant mass of the final 
$\eta^{\prime} p$ system and $q_{\pi}$ and $q_{\eta^{\prime}}$
are the momenta of the corresponding particles in the center-of-mass.
Since the $\pi N \rightarrow \eta^\prime N$ cross sections are 
known from the experimental data \cite{LB} the 
amplitude $|M_{\pi N \rightarrow 
\eta^\prime N}|$ can be extracted from the data using (\ref{amp2}). 
For the $\pi^+ n \to \eta^{\prime} p$
and $\pi^- p \to \eta^{\prime} n$ reactions this amplitude is shown in 
Fig.~\ref{fig1}; here the solid line corresponds to 
the average amplitude that
will be used in our following calculations. 
Note that $\sqrt{2}M_{\pi^0 p \to \eta^{\prime} p }$=
$M_{\pi^- p \to \eta^{\prime} n }$= 
$M_{\pi^+ n \to \eta^{\prime} p }$. For excess 
energies $\epsilon \leq$ 10 MeV
we need the $\pi N \to \eta^{\prime} N$ amplitude 
in the range $m_N+m_{\eta^{\prime}} \le \sqrt{s_1} \le
m_N+m_{\eta^{\prime}}+\epsilon $  
where $|M_{\pi N \to \eta^{\prime} N }|$ is almost constant.

\begin{figure}
\psfig{file=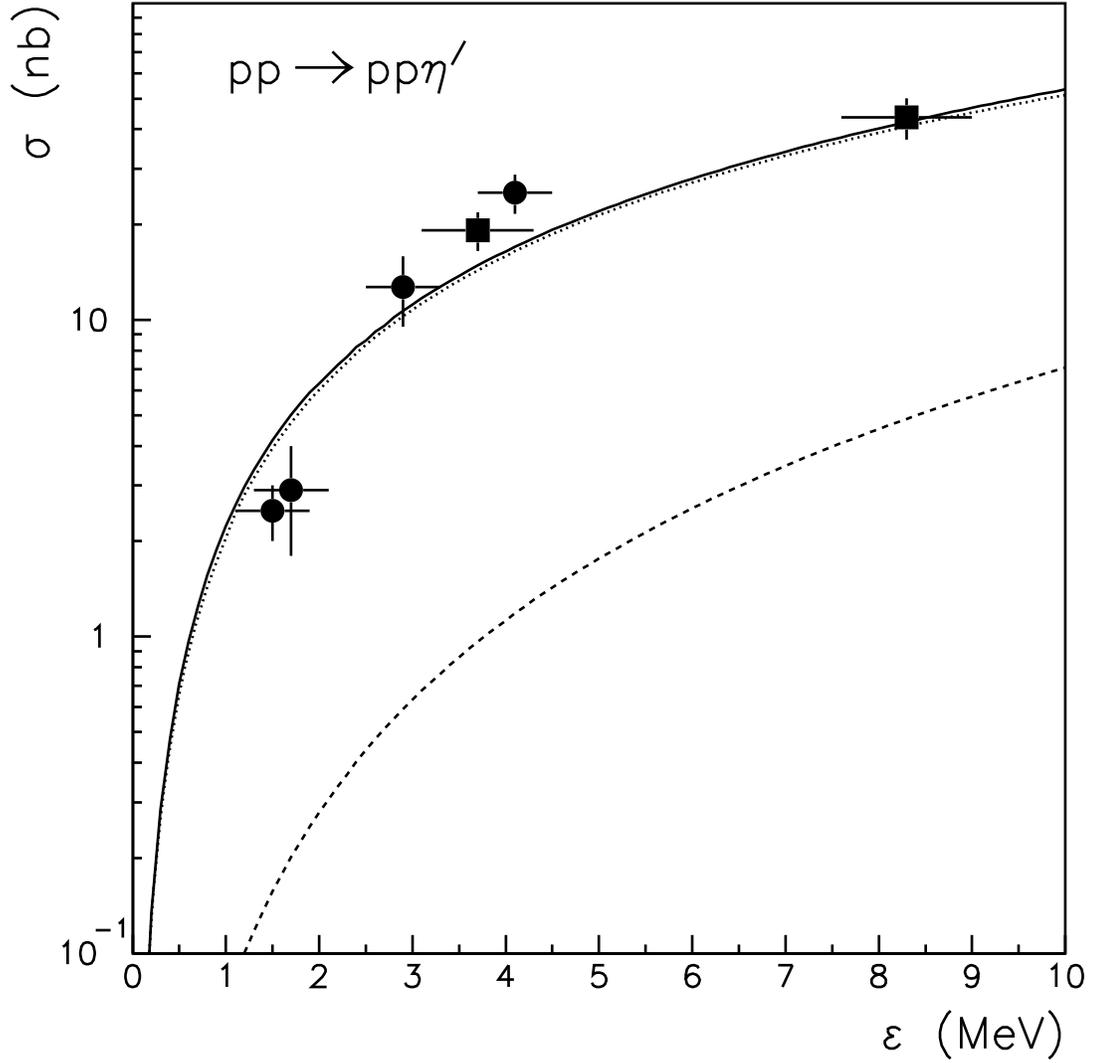,width=16cm}
\caption[]{The $pp \to pp \eta^{\prime}$ 
cross section as a function of the excess energy $\epsilon $. 
The experimental data are from Ref.~\protect\cite{Moskal} (circles) 
and Ref.~\protect\cite{Hibou} (squares). 
The dashed line corresponds to the pion-exchange calculation without FSI.
The solid and dotted lines are obtained including FSI according
to (\protect\ref{Jost}) and (\protect\ref{FW}), respectively.}
\label{fig2}
\end{figure}

The $pp \to pp \eta^{\prime}$ cross section then can be
obtained by integrating
\begin{equation}
\label{xsect}
\frac{d^2 \sigma }{dt \ ds_1} =\frac{1}
{2^9 \pi^3 q_a^2 s} \frac{q_{\eta^{\prime}}} {\sqrt{s_1}} \
|M_{prod} - exch.|^2 ,
\end{equation}
where $s$ is the squared invariant mass of the
colliding protons and $q_a$ is the momentum of the
incident proton in their center-of-mass. In~(\ref{xsect})
the exchange term is given by interchanging the initial proton 
momenta. The $pp \to pp \eta^{\prime}$ cross section 
calculated within  the model described above  is shown in 
Fig.~\ref{fig2} by the dashed line and substantially 
underestimates the experimental data from 
Refs.~\cite{Moskal,Hibou}. 

Following our analysis on the near threshold $K^+$-meson 
production~\cite{SiCa1} the latter discrepancy might be entirely 
due to FSI. Within the Watson-Migdal approximation~\cite{Watson,Migdal}
the total reaction amplitude can be factorized in terms of the
production amplitude~(\ref{amp1}) and the FSI amplitude.
As was shown by F\"aldt and Wilkin~\cite{Wilkin} (see 
also~\cite{GellMann}) the FSI might be 
introduced by multiplying the cross section~(\ref{xsect}) by the 
correction factor
\begin{equation}
\label{FW}
FW(\epsilon ) = {\left\lbrack
\frac{2\beta}{\alpha + \sqrt{{\alpha}^2+\epsilon m_N}} 
\right\rbrack}^2 ,
\end{equation}
where the parameters $\alpha$ and $\beta$ are related to the
scattering length $a$ and the effective range $r_0$ for the
S-wave $pp$ scattering as
\begin{equation}
\label{range}
a=\frac{\alpha + \beta}{\alpha \beta}, \hspace{2cm}
r_0= \frac{2}{\alpha + \beta}.
\end{equation}

\begin{figure}
\psfig{file=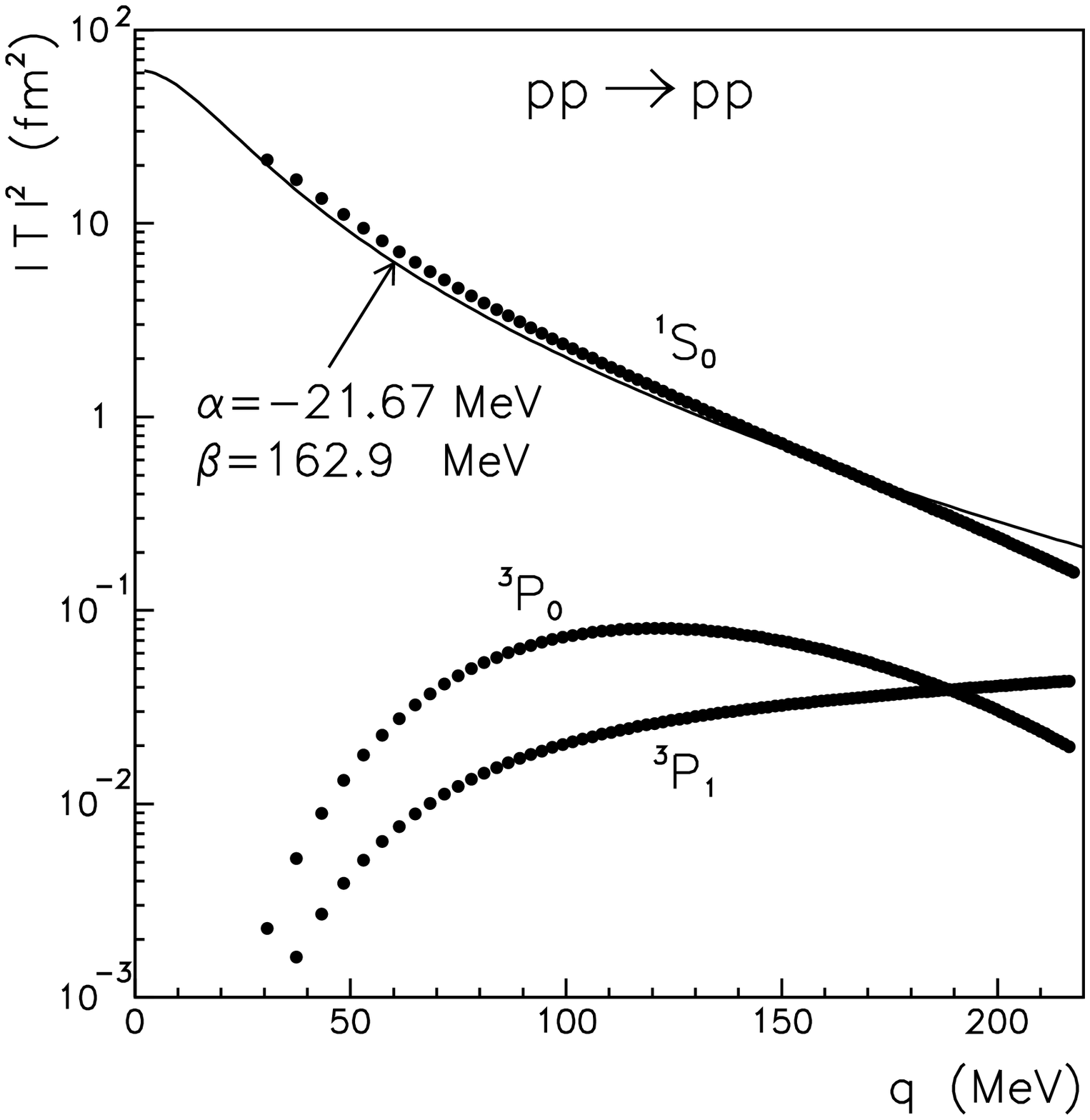,width=16cm}
\caption[]{The squared $pp$ scattering amplitude as
function of the relative momentum of the final protons.
Full dots show the results for the $^1S_0$, $^3P_0$ and
$^3P_1$ partial waves calculated with the Nijmegen-93
model~\protect\cite{Nijmegen}, while the line is the
effective range approximation with parameters as shown 
in the figure.}
\label{fig3}
\end{figure}

Furthermore, within the effective range expansion 
the S-wave scattering amplitude $T(q)$ and the phase shift 
$\delta $ are given
\begin{equation}
\label{scat}
T(q) = \frac{1}{q \, cot\delta - iq} = \left( -\frac{1}{a}+
\frac{r_0 q^2}{2}-iq \right)^{-1}.
\end{equation}
The squared $^1S_0$ $pp$ scattering amplitude $|T|^2$ calculated 
with the phase shift from the Nijmegen-93 partial wave 
analysis~\cite{Nijmegen} is shown in Fig.~\ref{fig3} in comparison
to the effective range approximation with parameters
$\alpha =-21.67$~MeV and $\beta =162.9$~MeV. Fig.~\ref{fig3} also
shows the contribution from $^3P_0$ and $^3P_1$ partial waves
and illustrates that the FSI is dominated by  S-wave proton-proton
scattering for relative momenta of the final protons below
$\simeq 100$~MeV/c. 

The result  obtained with the prescription~(\ref{FW})
is shown by the dotted line in Fig.~\ref{fig2} and provides a 
reasonable  description of the data.

Another way to account for the FSI between the protons is to
multiply the production amplitude~(\ref{amp1}) by the inverse 
S-wave Jost function~\cite{Taylor}
\begin{equation}
\label{Jost}
J_0(q) = \frac{q-i\alpha}{q+i\beta},
\end{equation}
where $q$ is the relative momentum of the final protons.
The solid line in Fig.~\ref{fig2} shows our calculation with FSI 
according to (\ref{Jost}) which reasonably reproduces the data
and can be compared with prescription~(\ref{FW}). 

Note, as  discussed in~\cite{SiCa1}, that the Jost function approaches 
unity for large $q$ while the Watson factor,
i.e. the use of the scattering amplitude itself, 
is reasonable only at energies close to the reaction threshold.   
Actually at energies close to the reaction threshold both models for FSI
corrections (\ref{Jost}) and (\ref{FW}) give a factor $\beta^2/\alpha^2$ 
for the production cross section.

We also note that the difference between the $pp$ and $pn$ 
FSI~\cite{Laget},
which might explain the large ratio of the 
$pp\to pp \eta $ and $pn \to pn \eta$ cross sections near threshold,
will lead to a comparable ratio of the 
$pp\to pp \eta^\prime$ and $pn \to pn \eta^\prime$ cross sections for low
excess energies. 

Since the one-pion-exchange model with the inclusion of 
the interaction between
the final protons reproduces the experimental data on
near-threshold ${\eta}^{\prime}$ production from $pp$
collisions, we conclude that other exchange currents 
in the primary production amplitude either 
play no role or cancel each other to a large extend.

\vspace{1cm}
We appreciate valuable  discussions with
U. Mosel as well as the communication with C. Wilkin.

\newpage

\newpage
\newpage


\begin{thebibliography}{99}
\bibitem{Moskal}
        Moskal, P.,  Balewski, J.T., Budzanowski, A., Dombrowski, H.,
        Goodman C., et al., Phys. Rev. Lett. {\textbf 80}, 3202 (1998);
\bibitem{Hibou}
        Hibou, F., Bing, O., Boivin, M., Courtat, P., F\"aldt, G.,
        et al., nucl-ex/9802002.
\bibitem{Germond} 
        Germond, J.-F., Wilkin, C., Nucl. Phys. A518, 308 (1990).
\bibitem{SiCa}
        Sibirtsev, A., Cassing, W., Preprint IFJ-1787-PH,
        nucl-th/980219.
\bibitem{Swart}
        De Swart, J.J., Rentmeester, M.C.M., Timmermans, R.G.E.,
        nucl-th/9802084.
\bibitem{Machleid}
        Machleid, R., Holinde, K., Elster, Ch., 
        Phys. Rep.  149, 1 (1987).
\bibitem{LB}
        Landolt-B{\"{o}}rnstein, \textit{New Series I/12 } 
        (ed. H. Schopper, Springer-Verlag, 1988). 
\bibitem{SiCa1}
        Sibirtsev, A., Cassing, W., nucl-th/9802025.
\bibitem{Watson}
        Watson, K.M., Phys. Rev. {\textbf 88}, 1163 (1952).
\bibitem{Migdal}
        Migdal, A.B., JETP {\textbf 1}, 2 (1955).
\bibitem{Wilkin}
        F\"aldt, G., Wilkin, C., 
        Phys. Lett. {\textbf B 382}, 209 (1996);
        Z. Phys. {\textbf A 357}, 241 (1997).
\bibitem{GellMann}
        Gell-Mann, M., Watson, K.M., Ann. Rev. Nucl. Sci.
        {\textbf 4}, 219 (1954).
\bibitem{Nijmegen}
        Stoks, V.G.J., Klomp, R.A.M., Rentmeester, M.C.M., de Swart,
        J.J., Phys. Rev. {\textbf C 48}, 792 (1993).
\bibitem{Taylor}
        Taylor, J.R., \textit{Scattering Theory} 
        (Willey, New York, 1972).
\bibitem{Laget}
        Laget, J.M., Wellers, F., Lecolley, J.F.,
        Phys. Lett. {\textbf B 257}, 254 (1991).
\end{thebibliography}
\end{document}